\documentstyle[aps,epsf]{revtex}
\begin{document}

\title
{Ab initio calculation of transport properties of metal-C$_{60}$-metal junctions}

\author{Nikolai Sergueev, Dan Roubtsov, and Hong Guo}
\address{Center for the Physics of Materials and Department of Physics,
McGill University, Montreal, P.Q., Canada H3A\,2T8}

\maketitle


\begin{abstract}
By carrying out density functional theory (DFT) analysis within the Keldysh
non-equilibrium Green's function (NEGF) framework, we investigate
the quantum transport properties of Au-C$_{60}$-Au molecular junctions from the first principles.  
We briefly review the NEGF-DFT formalism and present
some of our data. We found that at small electrode separations ($\simeq 10-13$\,{\AA}), 
the Au-C$_{60}$-Au junctions show metallic behaviour with $G\simeq (2e^{2}/h)$
and $I \simeq 1-3\,\mu$A  in the linear regime ($\vert V_{\rm bias}\vert < 0.3$\,V).    
The physical mechanism is the resonance tunnelling through partially occupied states 
originated from the lowest unoccupied molecular orbital (LUMO) of the free C$_{60}$. 
We also found that the charge transfer from the Au electrodes to the C$_{60}$
molecule can be controlled by gate potential.
\end{abstract}


\section{Introduction}
\label{introduction} 
Using molecules as functional units for electronic device 
application \cite{aviram} is an interesting perspective and a possible goal 
of nano-electronics. Work in this field has clearly demonstrated that many 
of the important molecular device characteristics relate specifically
to a strong coupling between the {\it atomic} and the {\it electronic}
degrees of freedom, for a popular introduction see reference \cite{PhToday}. However, from a theoretical point of view, the accurate 
prediction of the properties of atomic and molecular scale devices -- 
including the true I-V curves with as few adjustable parameters as 
possible -- still represents a formidable challenge \cite{Damle2001,Ventra2000,Palacios2001} despite the advances 
and wide-spread application of large scale {\it ab initio} modeling based on 
the density functional theory (DFT) over the last two decades 

Recall that most of the previous DFT-based {\it ab initio} 
simulations \cite{payne1992,ordejon1996} solve only two kinds of problems: 
(i) finite systems such as 
isolated molecules and clusters,  as in quantum chemistry; (ii) periodic systems consisting 
of super-cells, as in solid state physics. However, a molecular electronic 
device is neither finite nor periodic.  Typically,
it has open boundaries which 
are connected to long and different electrodes extending to electron 
reservoirs far away, and the external bias potentials are applied to these reservoirs.
In other 
words, calculations of finite or periodic systems do not include the correct 
boundary conditions for a nano-scale device in the quantum transport regime. Therefore, a new formalism for
electronic analysis is required to carry out the first principle transport
modeling for molecular electronics.

In this article, we briefly outline a formalism that combines the density functional theory with the
Keldysh non-equilibrium Green's functions (NEGF) so that non-equilibrium properties of 
the quantum transport regime can be predicted from atomistic approach
without any phenomenological parameters \cite{taylor2001a,brandbyge2002}.  
To show how it works, we apply our NEGF-DFT approach to investigate a C$_{60}$ molecule 
connected to two Au electrodes. The rest of the article is organized 
as follows: in section \ref{theory}, we present the NEGF-DFT formalism
and, in section \ref{results}, we present the transport properties of the 
Au-C$_{60}$-Au tunnel junctions and compare them with the properties of the Al-C$_{60}$-Al ones.  
Section \ref{summary} is reserved for a 
short summary.  

\begin{figure}
\begin{center}
\leavevmode
\epsfxsize = 8.8cm
\epsffile{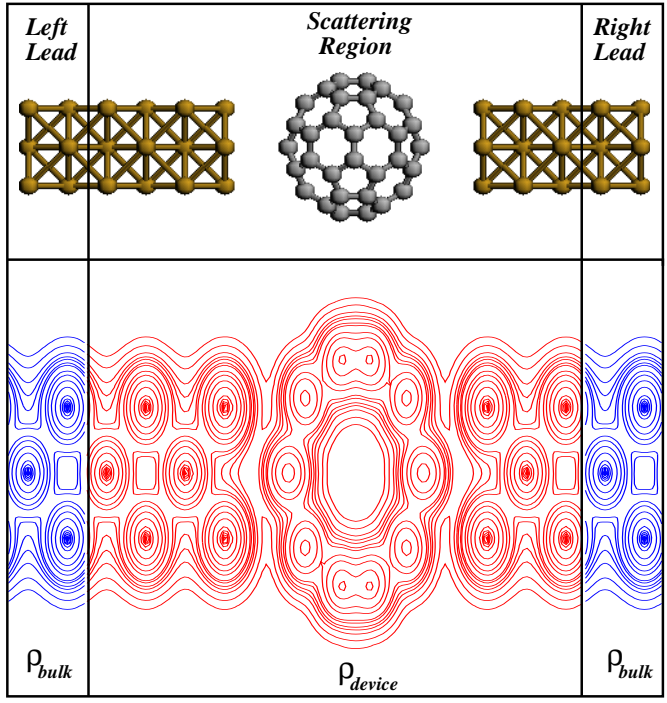}
\end{center}
\caption{Schematic plot of an Au-C$_{60}$-Au molecular
tunnel junction. The Au electrodes consist of repeated unit cells
extending to $z=\pm\infty$ of the horizontal axis $\vec{z}$ and its surface is oriented along the
(100) direction. The lower panel shows the calculated charge
density at equilibrium ($\mu_{l}=\mu_{r}$). Note that perfect matching is obtained across
the boundaries between the leads and the central scattering
region.} 
\label{device}
\end{figure}

\section{NEGF-DFT formalism}
\label{theory}

The simplest model of a molecular device is schematically shown in 
figure \ref{device} where a C$_{60}$ molecule is contacted by two semi-infinite 
Au electrodes which extend to $z=\pm \infty$ where bias voltage is applied.  
To calculate the electronic states of such a device, two problems should be
solved. First, the infinitely large problem (due to the electrodes) must be
reduced to something manageable on a computer, {\it i.e.}, one has to solve an open
boundary problem. Second, one has to find the charge density $\rho({\bf r})$ of the molecule and electrodes
provided there is a bias voltage across the device, and this is a non-equilibrium problem.
From the computational point of view, it is convenient to divide our system
into three parts: a scattering region with some portion of electrodes
and two metal electrodes extending to $z=\pm \infty$, as shown by the 
vertical lines in figure \ref{device}.

To reduce the infinitely large problem to that defined inside the scattering 
region (see figure \ref{device}), we notice that the effective Kohn-Sham (KS)
potential $V_{\rm eff}[\rho({\bf r})]$
deep inside the left or right lead is very close to the corresponding 
bulk KS potential. This fact makes the boundary conditions to be written in 
the following form \cite{taylor2001a}:
\begin{equation}
V_{\rm eff}({\bf r})=\left \{
\begin{array}{llll}
&V_{l,\,{\rm eff}}({\bf r})=V_{l,\,{\rm bulk}}({\bf r}),&\,\,\,z<z_l,\\
&V_{c,\,{\rm eff}}({\bf r}),&\,\,\,z_l<z<z_r,\\
&V_{r,\,{\rm eff}}({\bf r})=V_{r,\,{\rm bulk}}({\bf r}),&\,\,\,z>z_r,
\end{array}
\right. \label{eq-potential}
\end{equation}
where the planes $z=z_{l\,(r)}$ are the left (right) limits of the
scattering region (see figure \ref{device}),
and $V_{l,\,{\rm bulk}}({\bf r})$ and $V_{r,\,{\rm bulk}}({\bf r})$ are known.
In practice, within the DFT,
we only need to match the Hartree potential $V_{\rm H}^{}[\rho({\bf r})]$ at the 
boundaries. This can be accomplished by solving the Poisson equation on $U_{\rm H}^{}({\bf r})$
by use of a 3D real space grid and boundary conditions at $z=z_{l\,(r)}$, and \,
$V_{\rm H}^{}({\bf r})=eU_{\rm H}^{}({\bf r})$. 
Once the Poisson equation is solved, $V_{\rm H}^{}({\bf r})$ is matched, 
and by use of the DFT we can show that $V_{{\rm eff}}({\bf r})$ is also matched.
Thus, equation (1) is valid.

In addition, the 3D real space Poisson solver allows us to deal with any gate
potentials, $V_{g}$. In our model, they provide just additional boundary conditions for the 
electrostatics, so they change $U_{\rm H}^{}({\bf r})$.  
Furthermore, once $V_{\rm eff}({\bf r})$ is matched across
the boundaries, the charge density $\rho({\bf r})$ is automatically matched
at the boundaries.  The lower panel of figure \ref{device} shows the
charge density: although the densities in the leads and in the scattering
region are calculated separately, they match perfectly
across the boundaries.

The above ``screening approximation'' \cite{taylor2001a} allows us to reduce
the infinitely large Hamiltonian of the system to the effective finite one defined inside the scattering region.
In this approximation, we neglected any influence the scattering region may 
give to the leads, but if the portion of the leads included inside the 
scattering region is long enough, such an approximation is well controlled.  On the
other hand, the semi-infinite leads do contribute to the scattering region,
and this contribution is included through the self-energies $\Sigma_{l\,(r)}(E)$ in the Green's function $G(E)$
of the scattering region, for example, 
\begin{equation}
G^{R}(E)=\bigl(E-H_{0}-V_{\rm ps}-V_{\rm H}^{} - V_{\rm xc} -\Sigma_{l}^{R}-\Sigma_{r}^{R} \bigr)^{-1},
\label{G_ret}
\end{equation}
\begin{equation}
V_{\rm H}^{}=eU_{\rm H}^{},\,\,\,\,\,\Delta U_{\rm H}^{}({\bf r})=-4\pi\,\rho({\bf r}) .
\label{Poisson}
\end{equation}
Here, $H_{0}$ is the kinetic energy of a valent electron, and the atomic cores of
the scattering region  are fixed in space and described by pseudo-potentials, so we have 
$V_{\rm ps}$,
and $V_{\rm eff}[\,\rho\,]= V_{\rm H}^{}[\,\rho\,] + V_{\rm xc}[\,\rho\,]$ 
is the effective KS
potential obtained within the DFT. 
For example, the exchange correlation potential $V_{\rm xc}$ is taken in a simple 
local form  following the polynomial formula suggested in  reference \cite{God1},  
and the (retarded) self-energies due to the semi-infinite left and right electrodes are calculated following 
an iterative technique described in  references \cite{taylor2001a,Sun}. 

However, to calculate the charge density $\rho({\bf r})$ 
away from equilibrium due to the  bias 
voltage $V_{b}$, we have to apply the NEGF method \cite{jauho_book}. The density matrix of the scattering region
is calculated 
from the so-called lesser Green's function $G^<(E)$ of the scattering region  as follows:
\begin{equation}
\hat{\rho}= \frac{1}{2 \pi i }\int dE \,G^{<}(E),\,\,\,\mbox{where}\,\,\,
G^<(E)= G^R(E)\,\Sigma^{<}(E)\,G^A(E).  
\label{eq-noneq-green}
\end{equation}
Here, $G^{R\,(A)}(E)$ is the retarded (advanced) Green's function of the scattering region. 
Note that $G^{<}(E)$ is defined through the Keldysh equation \cite{jauho_book}. 
The lesser self-energy $\Sigma^{<}[f_l,\,f_r]$ can be evaluated within the mean field theory
as follows:
\begin{equation}
\Sigma^{<}(E)=i\,f_l(E,\,\mu_{l})\,\Gamma_{l}(E)+ i\,f_r(E,\,\mu_{r})\,\Gamma_{r}(E),
\end{equation}
where $\Gamma_{l\,(r)}=i\,(\Sigma_{l\,(r)}^{R}-\Sigma_{l\,(r)}^{A})$ and  
$\Sigma_{l\,(r)}^{R\,(A)}$ is the retarded (advanced) self-energy of the left
(right) electrode, and $f_{l\,(r)}(E)$ are the 
corresponding Fermi distribution functions.
Note that  $\Sigma^<[f_l,\,f_r]$ is more than a simple Fermi distribution: a fact reflecting 
the non-equilibrium nature of the problem, $\mu_{l}\ne \mu_{r}$.  

In our numerical procedure, we use a LCAO  minimal {\it s,\,p,\,d} atomic basis 
set (a real space fireball linear combination of atomic orbitals \cite{ordejon1996})
to expand the electron wavefunction and define the effective Hamiltonian matrix $F$:  
$$
H_{0}+V_{\rm ps}+V_{\rm H}^{}+V_{\rm xc}\,\rightarrow\,F_{jk}[\,\rho({\bf r})\,] 
$$
Here, to  calculate $V_{{\rm ps,}\,jk}$, the atomic cores are described 
by standard norm conserving nonlocal 
pseudo-potentials \cite{bachelet1982}.  
We evaluate $G^{R}$ 
by direct matrix inversion, $G^{A}(E)=G^{R\,\dag}(E)$, and  we construct $\rho({\bf r})$ from numerical 
evaluation of the integral in equation (\ref{eq-noneq-green}) using a suitable contour in 
the complex plane $E$ \cite{taylor2001a}.
 Once $\rho({\bf r})$ is obtained, we evaluate the KS potential 
and iterate the above procedure, equations (\ref{G_ret}),
(\ref{Poisson}), and (\ref{eq-noneq-green}), untill numerical convergence is reached.

Thus, we construct the self-consistent $\rho({\bf r})$ by the NEGF technique (it takes care of the
non-equilibrium statistics) and we calculate the Hartree potential by
solving the 3D Poission equation directly (equation (\ref{Poisson}) includes all the external 
fields as the electrostatic boundary conditions). As a result, our NEGF-DFT formalism
is able to solve charge transport problems in addition to the 
conventional electronic structure calculations: we know both the density of states ${\rm DOS}(E)$ and the density
matrix $G^{<}({\bf r},\,{\bf r}',\,E)/2\pi i$ of the scattering region.
Moreover, we can easily calculate the electron charge 
on the molecule at nozero $V_{b}$ and $V_{g}$. Of course, the density matrix $\hat{\rho}$
is normalized to the total number of valent electrons in the scattering region, 
$$
{\rm Tr}\,(\hat{\rho}\,S)=N_{\rm e}={\rm const},
$$ 
where $S$ is the overlap matrix of the atomic orbitals we use.
Then, one can introduce the  (average) number of electrons in the molecule by calculating this trace over the molecular orbitals 
belonging to the molecule only: 
$$
{\rm Tr}_{\rm mol}\,(\hat{\rho}\,S)=N_{\rm e,\,mol}(V_{b},\,V_{g}).
$$ 
The difference between the calculated $N_{\rm e,\,mol}(V_{b},\,V_{g})$ and  $N_{\rm e,\,mol}^{(0)}$ of a free molecule 
can called an (average) excess charge on the molecule in the units of $e$,
$$
Q=N_{\rm e,\,mol}(V_{b},\,V_{g})-N_{\rm e,\,mol}^{(0)}.
$$

\section{Au-C$_{60}$-Au molecular tunnel junction} 
\label{results}

In this section we report our analysis on the transport properties
of an Au-C$_{60}$-Au molecular tunnel junction calculated by use of
our NEGF-DFT electronic package as discussed above. The device structure 
is shown in the upper panel of figure \ref{device}. So far a considerable amount 
of theoretical and experimental effort has been devoted to investigate 
transport properties of C$_{60}$ and other fullerene 
molecules \cite{taylor2001,park,alavi02}.
The device we present consists of the fullerene molecule C$_{60}$ fixed in the middle  
of two gold electrodes. Here, we consider two distances between the electrodes, $11.7$\,{\AA} and $13.7$\,{\AA}.
This correspondes to the minimum distance between an Au atom and C atom of $2.3$\,{\AA} and  $3.3$\,{\AA}, respectively.
Each electrode consists of repeated unit cells with nine Au atoms in the (100) 
direction and extended to infinity. Note that we do not expect that the C$_{60}$ molecule  
can be found in the Coulomb blockade 
regime in such junctions if the electrode separation is small, say  $9-13$\,{\AA}. 
The number of electrons in the molecule is not a good quantum number 
if the interaction with electrodes is strong.   

The current-voltage (I-V) characteristics is calculated as follows:
\begin{equation}
I=\frac{2e^2}{h}\int T(E,\,V_b)\,[f_l(E)-f_r(E)]\,dE \,\,, 
\label{current1}
\end{equation}
where $T(E,\,V_b)$ is the transmission coefficient at the energy $E$ and bias 
voltage $V_b$. The Fermi distribution functions $f_{l\,(r)}(E)$ limit 
the integration in  equation (\ref{current1}) to the small energy range $e V_b$ at the Fermi level 
of the electrodes (in the zero temperature limit). 
Transmission coefficient $T(E,V_b)$ is calculated 
by use of the Green's functions \cite{datta_book} which we have already obtained as a result of 
the NEGF-DFT self-consistent iterations, so we write 
\begin{equation}
T(E,\,V_b)={\rm Tr}\,\bigl[\, \Gamma_l(E)\,G^{R}(E)\,\Gamma_{r}(E)\,G^{R\,\dag}(E)\,\bigr].
\label{trans1}
\end{equation}
Here, the matrix ${\Gamma}_{l\,(r)}(E)$ is related to the line-widths resulting 
from the coupling of the scattering region to the left (right) lead. It is 
evaluated in terms of the corresponding self-energy matrices $\Sigma_{r\,(l)}$ as
follows:
\begin{equation}
\Gamma_{l\,(r)}(E)=i\,\bigl(\,\Sigma_{l\,(r)}^{R}(E)-\Sigma_{l\,(r)}^{R\,\dag}(E)\,\bigr).
\end{equation}
The retarded Green's function   of the scattering region $G^R(E)$  
is given as the following matrix:
\begin{equation}
G^{R}(E)=\bigl(\,ES- F[\,\rho\,] - \Sigma^{R}(E)\,\bigr)^{-1}, 
\label{green1}
\end{equation}
where $F[\,\rho\,]$ is the effective Hamiltonian of the scattering region and 
$\Sigma^{R}=\Sigma_l^R+ \Sigma_r^R$ is the total self-energy matrix.
The density of state function is calculated from the density of state matrix 
$A(E)=i\,\bigl(G^{R}(E)-G^{R\,\dag}(E)\bigr)$ as follows:
$$
{\rm DOS}(E)={\rm Tr}\,\bigl(A(E)\,S\bigr)/2\pi.
$$

\begin{figure}
\begin{center}
\leavevmode
\epsfxsize = 9.45cm
\epsffile{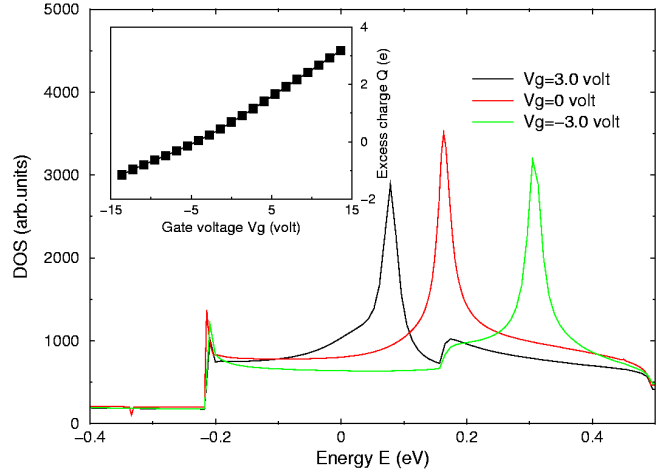}
\end{center}
\caption{Density of states of the Au-C$_{60}$-Au junction and excess charge of the C$_{60}$ molecule are presented for the electrode
separation of 11.7\,{\AA} and $V_{b}=0$. At the zero gate, the excess charge is $\approx 0.7\,e$. Note that the C$_{60}$ molecule 
is not in the Coulomb blockade regime (the broadening is large).}
\label{DOS1} 
\end{figure}

\begin{figure}
\begin{center}
\leavevmode
\epsfxsize = 9.45cm
\epsffile{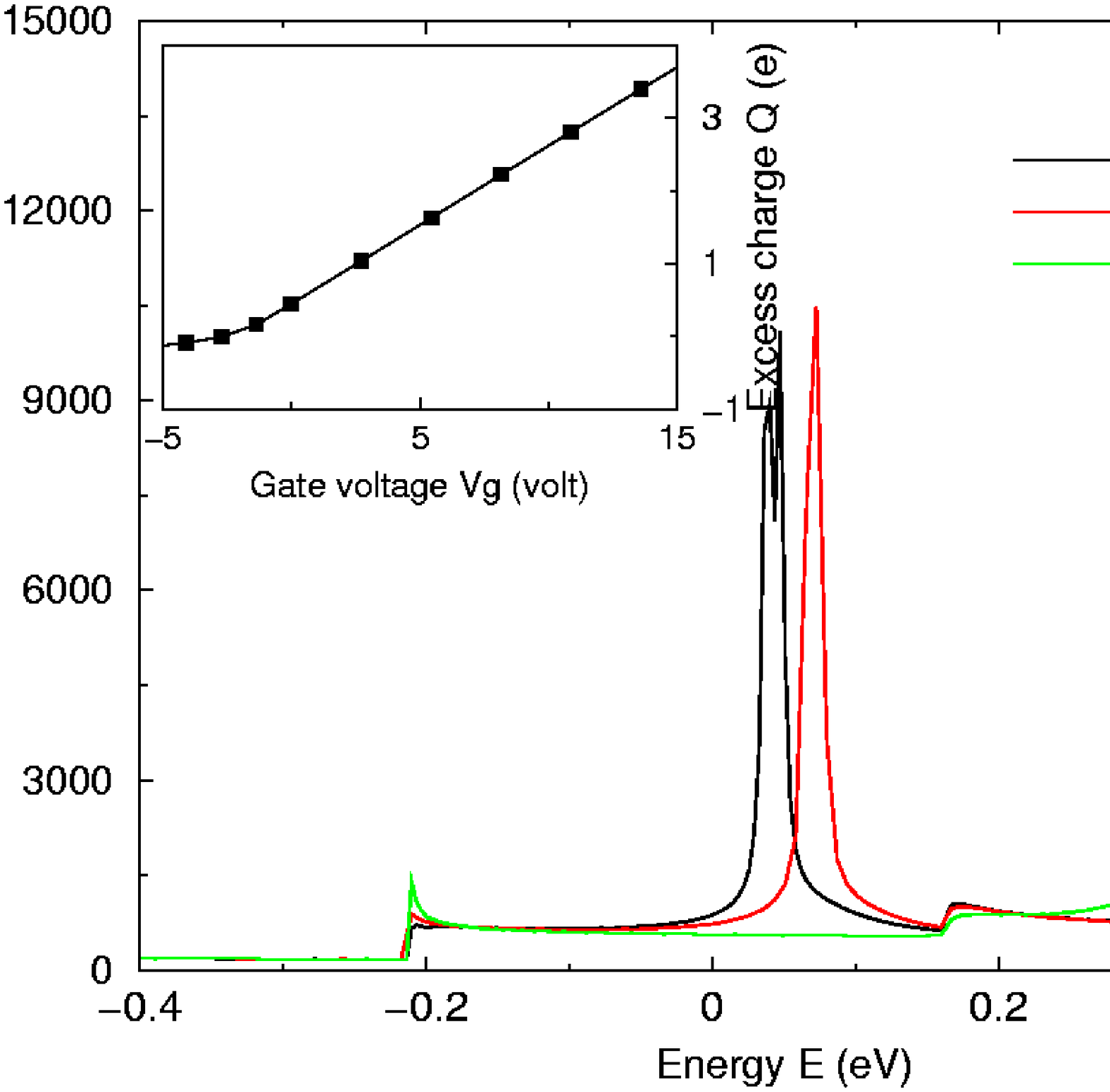}
\end{center}
\caption{Density of states of the Au-C$_{60}$-Au junction and  excess charge of the C$_{60}$ molecule are presented for the electrode
separation of 13.7\,{\AA} and $V_{b}=0$. At the zero gate, the excess charge is $\approx 0.5\,e$. 
Note that the C$_{60}$ molecule 
is not in the Coulomb blockade regime yet.}
\label{DOS2}
\end{figure}

For the Au-C$_{60}$-Au junctions at $V_{b}=0$, 
the density of states near the Fermi energy  is presented on figures \ref{DOS1} and \ref{DOS2} 
(the Fermi level is set to  $E_{\rm F}^{}=0$ in this article). Note that at positive gate voltages, $V_{g}\ge 0$, 
the energy levels originated from the LUMO of C$_{60}$ are partially occupied so they lie near $E=0$ and $Q>0$ (in the units of $e$). 
At negative gate voltages, $V_{g}< V_{g}^{*}<0$, 
the energy levels originated from the heighest occupied molecular orbital (HOMO) of C$_{60}$ begin to be depopulated, 
so the energy levels originated from the LUMO of C$_{60}$
are shifted far to the right from $E=0$ and $Q<0$ (in the units of $e$).

\newpage

\newpage
\begin{figure}
\begin{center}
\leavevmode
\epsfysize = 200pt
\epsffile{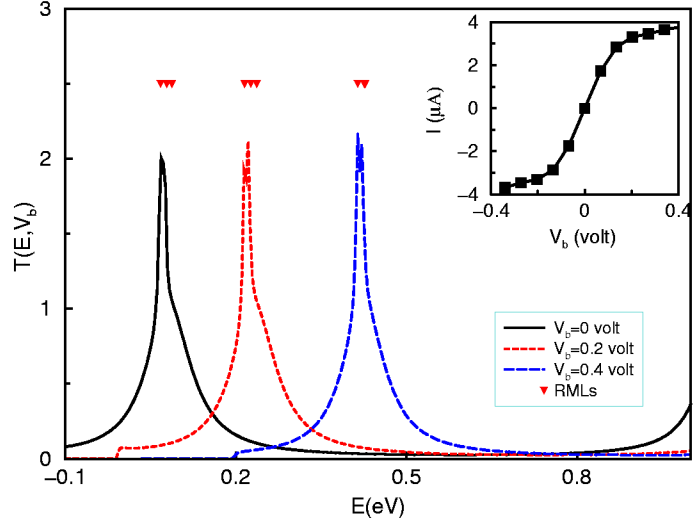}
\end{center}
\caption{ 
Transmission coefficient $T(E,\,V_b)$ versus electron
energy $E$ is presented for the electrode
separations of 11.7\,{\AA} and  three bias voltages $V_b$, $V_{g}=0$. The sharp transmission
peak, which results from a resonances of the fullerene\,+\,contacts
Hamiltonian,  
dominates the conductance as well as the
current-triggered dynamics.  Positions of 
the Renormalized Molecular Levels (RML) [15] of C$_{60}$ 
are depicted over the peaks of $T(E,\,V_b)$, and these levels are the LUMO-derived states.
The inset shows the calculated I-V
curve from which the metallic behaviour of the junction is evident.} 
\label{iv}
\end{figure}

The transmission coefficient obtained from equation (\ref{trans1}) is presented 
on  figure \ref{iv} as a function of the electron energy $E$ for three different 
bias voltages and the zero gate potential.  The sharp peak in $T(E,\,V_b)$ is the 
result of a resonance transmission through the 
LUMO of the C$_{60}$. At $V_{b}=0$, this resonance lies just above the 
Fermi level of the leads, see figures \ref{DOS1} and \ref{DOS2}. 
As the bias voltage is applied, the peak position shifts toward higher energies. 
If the system is absolutely symmetric and $V_{b}\ne 0$, we would expect the same voltage drop 
at the two electrode-molecule junctions. In our system, the atomic structures of 
$C_{60}$ facing the left and right electrodes are not the same, and this breaks the 
left-right symmetry. We therefore observe an asymmetric voltage drop at 
the two contacts. Consequently, the transmission peak position shifts by 
about $\sim 0.15$\,eV when $V_b$ is increased by $0.2$\,V, {\it i.e.}, the applied 
voltage drops more on one side than on the other side of the C$_{60}$. We also 
calculated the I-V curve for this device by use of equation (\ref{current1}); it is shown in 
the inset of figure \ref{iv}.

\begin{figure}
\begin{center}
\leavevmode
\epsfxsize = 9.45cm
\epsffile{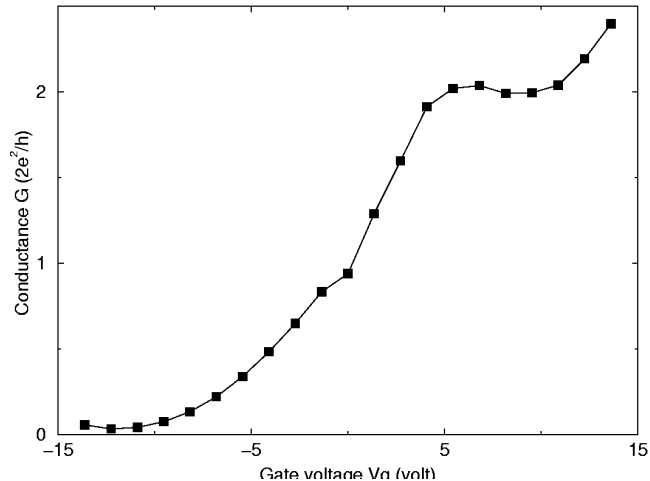}
\end{center}
\caption{Equilibrium conductance G is plotted as a function of the gate volage $V_{g}$ for the electrode
separations of 11.7\,{\AA}. Note that $G \approx G_{0}$ at $V_{g}=0$.}
\label{G1} 
\end{figure}
\begin{figure}
\begin{center}
\leavevmode
\epsfxsize = 9.45cm
\epsffile{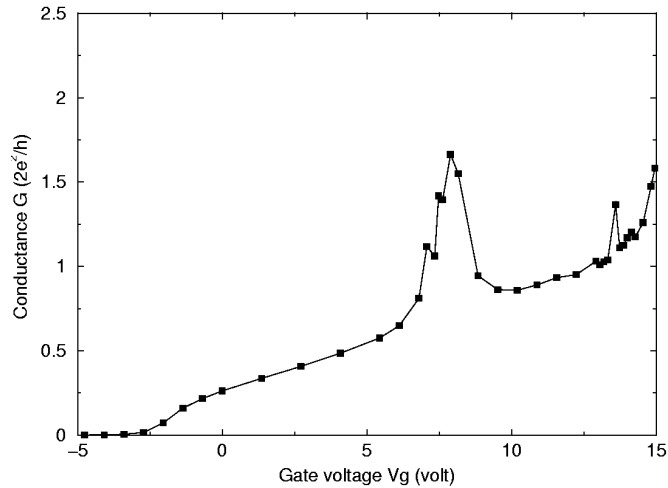}
\end{center}
\caption{Equilibrium conductance G is plotted as a function of the gate volage $V_{g}$ for the electrode
separations of 13.7\,{\AA}. Note that $G \approx 0.25\,G_{0}$ at $V_{g}=0$.}
\label{G2}
\end{figure}

We can easily calculate the equilibrium conductance $G$ of the junction, $G=dI/dV$.
In the linear regime, $-0.25\,{\rm V}<V_{b}<0.25$\,V, see figure \ref{iv}, we can use the following formula:
\begin{equation}
G=G_{0}\,T(E=E_{\rm F}^{},\, V_{b}=0)
\label{conduct}
\end{equation}   
(valid in the zero temperature limit).  Here, $G_{0}=2e^{2}/h$ is the quantum of conductance.  As the positive gate potential changes the occupancy of the 
LUMO-derived energy levels, we can expect the strong dependence of $G(V_{g})$
for different separations of the leads.
These results  are presented in figures \ref{G1} and  
\ref{G2}.

Similar metallic behaviour has been reported 
before \cite{taylor2001} for Al-C$_{60}$-Al junctions at the electrode separation of $9.3$\,{\AA} 
(it correspondes to the minimum distance between an Al atom and C atom of $\approx 1.1$\,{\AA}).  
Note that the work function of Al is $\approx 4.19$\,eV whereas the work function of Au is $\approx 5.3$\,eV.
At $V_{b}=V_{g}=0$, we have  $Q\approx 3\,e$, \,$G \approx 2.2\,G_{0}$
for the Al-C$_{60}$-Al junction \cite{taylor2001} and $Q\approx 0.7\,e$, \,$G \approx G_{0}$  for the Au-C$_{60}$-Au one 
(the electrode separation of 11.7\,{\AA}). 
In addition, for  
the Au-C$_{60}$-Au device,
the current in the linear (metallic) regime is (roughly) one order of value smaler than the current calculated for the Al-C$_{60}$-Al one. 

Here, the interesting physics is that despite the large HOMO-LUMO gap of a free $C_{60}$ 
($\simeq 1.8$\,eV \cite{cage}), we predict metallic transport characteristics for the
Au-C$_{60}$-Au devices, and they are qualitatively the same as the properties of the Al-C$_{60}$-Al devices.
We found that if the $C_{60}$ molecule is relatively well bonded 
to metallic leads, there is a strong charge transfer from the leads to the 
C$_{60}$ cage and it partially fills the (empty) LUMO of C$_{60}$. In other 
words, charge transfer from the electrodes to C$_{60}$ aligns the LUMO to the Fermi 
energy of the leads. As a result, we obtain a large resonance conductance and the metallic 
I-V curve.

\section{Summary}
\label{summary}

We have shown that the NEGF-DFT formalism is a powerful technique for modeling
charge transport properties of molecular electronic devices.  The novelty 
of this technique is in constructing electron charge density via 
nonequilibrium Green's functions and the very effective screening 
approximation. Since the entire algorithm is based on evaluating Green's 
function, the technique is intrinsically $O(N)$ \cite{God2}: this is because atoms far 
away from each other do not overlap so that the Hamiltonian matrix is  block-diagonal. This has tremendous computational advantage which 
we demonstrated by calculating the nonlinear I-V curve of the Au-C$_{60}$-Au 
molecular tunnel junction.

\section*{Acknowledgments}
We gratefully acknowledge financial support from Natural Science and
Engineering Research Council of Canada, le Fonds pour la Formation de
Chercheurs et l'Aide \`a la Recherche de la Province du Qu\'ebec, and
NanoQuebec. We gratefully acknowledge Jeremy Taylor and Brian 
Larade for their contributions throughout the work presented here.


\end{document}